\documentclass[11pt]{article}
\usepackage[dvips]{color}
\usepackage{epsfig}
\usepackage{amsmath}
\usepackage{graphicx}
\date{}
\begin{document}
\title{{\bf Bianchi type I, Schutz perfect fluid and evolutionary quantum cosmology}}
\author{Fatimah Tavakoli\thanks{%
e-mail: tavakoli09@gmail.com}\,\, and Babak Vakili\thanks{e-mail:
b.vakili@iauctb.ac.ir (corresponding author)}\\\\{\small {\it
Department of Physics, Central Tehran Branch, Islamic Azad
University, Tehran, Iran}}} \maketitle

\begin{abstract}
We study the classical and quantum cosmology of a universe in which
the matter content is a perfect fluid and the background geometry is
described by a Bianchi type I metric. To write the Hamiltonian of
the perfect fluid we use the Schutz representation, in terms of
which, after a particular gauge fixing, we are led to an
identification of a clock parameter which may play the role of time
for the corresponding dynamical system. In view of the classical
cosmology, it is shown that the evolution of the universe represents
a late time expansion coming from a big-bang singularity. We also
consider the issue of quantum cosmology in the framework of the
canonical Wheeler-DeWitt (WDW) equation. It is shown that the Schutz
formalism leads to the introduction of a momentum that enters
linearly into Hamiltonian. This means that the WDW equation takes
the form of a Schr\"{o}dinger equation for the quantum-mechanical
description of the model under consideration. We find the
eigenfunctions and with the use of them construct the closed form
expressions for the wave functions of the universe. By means of the
resulting wave function we evaluate the expectation values and
investigate the possibility of the avoidance of classical
singularities due to quantum effects. We also look at the problem
through Bohmian approach of quantum mechanics and while recovering
the quantum solutions, we deal with the reason of the singularity
avoidance by introducing quantum potential.

\vspace{5mm}\noindent\vspace{0.8mm}\newline Keywords:
Bianchi-I model; Quantum cosmology; Perfect fluid; Schutz formalism
\end{abstract}
\section{Introduction}
One of the most important questions in cosmology is that of the
initial conditions from which the universe began to evolve. As is
well known, standard theories of cosmology based on classical
general relativity do not provide an acceptable answer to this
question. The main reason is that almost all of these models suffer
from the existence of various types of singularities such as
big-bang, big-crunch, big-rip and so on. In the presence of any of
these singularities, some physical observables take values that are
physically unacceptable, which in turn means that their underlying
theory is not valid in the vicinity of singularities and thus can
not be applied. As to how the physical phenomena should be described
near these singularities, the dominant belief implies the
development of a concomitant and conducive quantum theory of
gravity. These efforts have begun with the works of DeWitt in
canonical quantum gravity \cite{DeW} and so far continue with the
more modern approaches such as string theory and loop quantum
gravity \cite{lqg}. Based on quantum gravity it would be useful to
describe the state of the universe near the classical singularities
within the framework of quantum cosmology. Depending on the theory
of quantum gravity the formalism of the corresponding quantum
cosmology is based on the canonical or loop quantization of the
minisuperspace variables and the evolution universe is described by
a wave function in this space \cite{Wilt}.

In this paper we deal with the issue of the classical and quantum
cosmology in the framework of the Bianchi type I model. Bianchi
models are the most well-known anisotropic and homogeneous
space-times whose classical and quantum solutions have been studied
in a number of works, see for example \cite{Ryan}. On the other
hand, an important ingredient in any model theory related to
cosmology is the choice of the matter field used to couple with
gravity and construct the energy-momentum tensor in Einstein field
equations. Inspired by the Weyl principle in cosmology, the most
widely used matter source has traditionally been the perfect fluid.
So, in the present study, we consider a perfect fluid as the matter
content of the universe and describe its evolution in the framework
of Schutz formalism \cite{Schutz}. In Schutz formulation of the
perfect fluid, its four-velocity can be expressed in terms of some
thermodynamical potentials. The advantage of using this
representation is that a canonical transformation can transform a
pair of the fluid's dynamical variable to another conjugate pair
$(T,p_T)$ in such a way that the momentum $p_T$ associated to the
variable $T$ appears linearly in the Hamiltonian. This means that
Schutz representation of perfect fluid can offer a time parameter in
terms of dynamical variables of the fluid. Therefore, when canonical
quantization we are facing with a Schr\"{o}dinger like equation in
which the wave function, in addition to the dependence on the
configuration space variables, has time dependence \cite{Time}. In
this way, it can be seen that by the Schutz formalism we may address
the problem of time in quantum cosmology.

In the following, in section 2, after a quick look at some main
properties of the Bianchi type I metric, we write the total
Hamiltonian based on the Einstein-Hilbert action coupled to a
perfect fluid in Schutz representation. Section 3 is devoted to the
classical Hamiltonian equations of motion and their solutions. Here,
in terms of the above mentioned time parameter, we shall obtain the
dynamical behavior of the cosmic scale factors that are the comoving
volume function of the universe and its anisotropic factors. We show
that the evolution of the universe represents a late time power law
expansion coming from a big-bang singularity. In section 4, we deal
with the quantization of the model. By the resulting wave function,
we compute the expectation values of the scale factors and show that
the evolution of the universe according to the quantum picture is
free of classical singularities but in good agreement with classical
dynamics in late times of cosmic evolution. Section 5 deals with the
Bohmian trajectories of the problem at hand, by means of which we
introduce the quantum potential. We will see that the repulsive
force associated to the quantum potential may be considered
responsible for the elimination of singularity. Finally, we
summarize the results in Section 6.

\section{The model}
In this section we make a brief overview of the most important
features of the Bianchi type I model and obtain its Lagrangian and
Hamiltonian in terms of the ADM variables in $1+3$ decomposition.
Bianchi type I universe is the simplest anisotropic generalization
of the flat FRW space-time whose metric reads as

\begin{equation}\label{A}
ds^2=-N^2(t)dt^2+a^2(t)dx^2+b^2(t)dy^2+c^2(t)dz^2.\end{equation} In
this metric $N(t)$ is the lapse function and there are three
functions $a(t)$, $b(t)$ and $c(t)$, to be determined by the
gravitational field equations, and are the scale factors of the
corresponding universe in the $x$, $y$ and $z$ directions
respectively. When writing the Lagrangian of this model, the scale
factors in different directions are considered as independent
variables. The Bianchi I metric as a cosmological setting is the
simplest homogeneous and anisotropic model, which becomes the flat
FRW metric provided that its scale factors are equal.

In general, Bianchi I space-time is a subset of the class A Bianchi
metrics with nine models. They are the most general homogeneous
dynamical solutions of the Einstein field equations which admit a
three-dimensional isometry group, i.e. their spatially homogeneous
sections are invariant under the action of a three-dimensional Lie
group. In order to transform the Lagrangian of the dynamical system
which corresponds to the Bianchi models to a more manageable form,
let us introduce the following change of variables

\begin{equation}\label{B}
a(t)=e^{u(t)+v(t)+\sqrt{3}w(t)},\hspace{0.5cm}b(t)=e^{u(t)+v(t)-\sqrt{3}w(t)},\hspace{0.5cm}c(t)=e^{u(t)-2v(t)}.\end{equation}
In the Misner notation \cite{Misner}, the metrics of all class A
Bianchi models can be written in terms of the above variables as

\begin{equation}\label{C}
ds^2=-N^2(t)dt^2+e^{2u(t)}e^{2\beta_{ij}(t)}dx^idx^j,
\end{equation}where $V(t)=e^{3u(t)}=abc$ is the comoving volume of the universe and $\beta_{ij}$ determines the
anisotropic parameters $v(t)$ and $w(t)$ as follows

\begin{equation}\label{D}
\beta_{ij}=\mbox{diag}\left(v+\sqrt{3}w,v-\sqrt{3}w,-2v\right).
\end{equation}
The action of such a structure may be written as (we work in units
where $c =\hbar=8\pi G=1$)

\begin{equation}\label{E}
{\cal S}=\frac{1}{2}\int_M d^4 x\sqrt{-g}{\cal R}+\int_{\partial
M}d^3x\sqrt{h}h_{ij}K^{ij}+\int_M d^4x\sqrt{-g}p,\end{equation}
where $g$ is the determinant and ${\cal R}$ is the Ricci scalar of
the space-time metric (\ref{A}). Also, $K^{ij}$ is the extrinsic
curvature (second fundamental form), which represents how much the
spatial space is curved in the way it sits in the space-time
manifold, and $h_{ij}$ is the induced metric over the
three-dimensional spatial hypersurface, which is the boundary
$\partial M$ of the four-dimensional manifold $M$. The boundary term
in (\ref{E}) will be canceled by the variation of the first term.
This is the reason for including a boundary term in the action of a
gravitational theory. That such a boundary term is needed is due to
the fact that ${\cal R}$, the gravitational Lagrangian density
contains second derivatives of the metric tensor, a nontypical
feature of field theories. The last term of (\ref{E}) denotes the
matter contribution to the total action where $p$ is the pressure of
perfect fluid which is linked to its energy density by the equation
of state (EoS)

\begin{equation}\label{F}
p=\omega \rho,
\end{equation}where $-1\leq\omega\leq 1$ is the EoS parameter.
In terms of the ADM variables, the gravitational part of the action
(\ref{E}) can be written as \cite{book1}

\begin{equation}\label{G}
{\cal S}_g=\frac{1}{2}\int dt d^3x{\cal L}=\frac{1}{2}\int dt d^3x
N\sqrt{h}\left(K_{ij}K^{ij}-K^2+R\right),\end{equation}where $h$ and
$R$ are the determinant and Ricci scalar of the spatial geometry
$h_{ij}$ respectively, and $K$ represents the trace of $K_{ij}$. The
extrinsic curvature is given by

\begin{equation}\label{H}
K_{ij}=\frac{1}{2N}\left(N_{i|j}+N_{j|i}-\frac{\partial
h_{ij}}{\partial t}\right),\end{equation}where $N_{i|j}$ represents
the covariant derivative with respect to $h_{ij}$. Using (\ref{C})
and (\ref{D}) we obtain the nonvanishing components of the extrinsic
curvature and its trace as follows

\begin{eqnarray}\label{I}
\left\{
\begin{array}{ll}
K_{11}=-\frac{1}{N}(\dot{u}+\dot{v}+\sqrt{3}\dot{w})e^{2(u+v+\sqrt{3}w)},\\\\
K_{22}=-\frac{1}{N}(\dot{u}+\dot{v}-\sqrt{3}\dot{w})e^{2(u+v-\sqrt{3}w)},\\\\
K_{33}= -\frac{1}{N}(\dot{u}-2\dot{v})e^{2(u-2v)},\\\\
K=-3\frac{\dot{u}}{N},
\end{array}
\right.
\end{eqnarray}
where a dot represents differentiation with respect to $t$. It is
easy to show for the Bianchi-I metric the Ricci scalar of the three
dimensional spatial geometry $h_{ij}$ vanishes, $R=0$. The
gravitational part of Lagrangian for the Bianchi-I model may now be
written by substituting the above results into action (\ref{G}),
giving
\begin{equation}\label{J}
{\cal
L}_g=\frac{3e^{3u}}{N}\left(-\dot{u}^2+\dot{v}^2+\dot{w}^2\right).
\end{equation}The momenta conjugate to the dynamical variables are given by
\begin{equation}\label{K}
p_u=\frac{\partial {\cal L}}{\partial
\dot{u}}=-\frac{6}{N}\dot{u}e^{3u},\hspace{.5cm}p_v=\frac{\partial
{\cal L}}{\partial
\dot{v}}=\frac{6}{N}\dot{v}e^{3u},\hspace{.5cm}p_w=\frac{\partial
{\cal L}}{\partial \dot{w}}=\frac{6}{N}\dot{w}e^{3u},
\end{equation}
leading to the following Hamiltonian
\begin{equation}\label{L}
{\cal H}_g=\frac{1}{12}Ne^{-3u}\left(-p_u^2+p_v^2+p_w^2\right).
\end{equation}
Now, let us deal with the matter field with which the action of the
model is augmented. The matter will come into play in a common way
and the total Hamiltonian can be made by adding the matter
Hamiltonian to the gravitational part (\ref{L}). According to
Schutz's representation for the perfect fluid, its Hamiltonian can
be viewed as (see \cite{Time} for details)

\begin{equation}\label{M}
{\cal H}_m=N\frac{P_T}{e^{3\omega u}},
\end{equation}where $T$ is a dynamical variable related to the thermodynamical parameters of the
perfect fluid and $P_T$ is its conjugate momentum. Finally, we are
in a position in which can write the total Hamiltonian $H={\cal
H}_g+{\cal H}_m$ as

\begin{equation}\label{N}
H=N\left[\frac{1}{12}e^{-3u}\left(-p_u^2+p_v^2+p_w^2\right)+P_T
e^{-3\omega u}\right].
\end{equation}
The setup for constructing the phase space and writing the
Lagrangian and Hamiltonian of the model is now complete. In the
following sections, we shall deal with classical and quantum
cosmologies which can be extracted from a theory with the above
mentioned Hamiltonian.

\section{Classically cosmological dynamics}
The classical dynamics is governed by the Hamiltonian equations,
that is
\begin{eqnarray}\label{O}
\left\{
\begin{array}{ll}
\dot{u}=\{u,H\}=-\frac{N p_u e^{-3 u}}{6},\\\\
\dot{p_u}=\{p_u,H\}=N \left[3 \omega  P_T e^{-3 \omega
u}+\frac{1}{4}
e^{-3 u} \left(-p_u^2+p_v^2+p_w^2\right)\right],\\\\
\dot{v}=\{v,H\}=\frac{N p_v e^{-3 u}}{6},\\\\
\dot{p_v}=\{p_v,H\}=0,\\\\
\dot{w}=\{w,H\}=\frac{N p_w e^{-3 u}}{6},\\\\
\dot{p_w}=\{p_w,H\}=0,\\\\
\dot{T}=\{T,H\}=N e^{-3 \omega
u},\\\\
\dot{P_T}=\{P_T,H\}=0.
\end{array}
\right.
\end{eqnarray}We also have the constraint equation $H=0$ which comes from the variation with respect to $N$. Up to this point
the cosmological model, in view of the concerning issue of time, has
been of course under-determined. Before trying to solve these
equations we must decide on a choice of time in the theory. The
under-determinacy problem at the classical level may be resolved by
using the gauge freedom via fixing the gauge. A glance at the above
equations shows that choosing the gauge $N=e^{3\omega u}$, we have

\begin{equation}\label{P}
N=e^{3\omega u}\Rightarrow \dot{T}=1\Rightarrow T=t,
\end{equation}which means that variable $T$ may play the role of time in the
model. Therefore, the classical equations of motion can be rewritten
in the gauge $N=e^{3\omega u}$ as follows

\begin{eqnarray}\label{Q}
\left\{
\begin{array}{ll}
\dot{u}=-\frac{1}{6}e^{-3(1-\omega)u}p_u,\\\\
\dot{p_u}=3\omega p_{0}+\frac{1}{4}e^{-3(1-\omega)u}\left(-p_u^2+p_{0v}^2+p_{0w}^2\right),\\\\
\dot{v}=\frac{1}{6}p_{0v}e^{-3(1-\omega)u},\\\\
\dot{w}=\frac{1}{6}p_{0w}e^{-3(1-\omega)u},
\end{array}
\right.
\end{eqnarray}in which we have set $p_v=p_{0v}=\mbox{const}$,
$p_w=p_{0w}=\mbox{const}$ and $P_T=p_0=\mbox{const}$. In addition
the Hamiltonian constraint $H=0$, yields

\begin{equation}\label{R}
\frac{1}{12}e^{-3u}\left(-p_u^2+p_v^2+p_w^2\right)+P_T e^{-3\omega
u}=0,
\end{equation}from which we get

\begin{equation}\label{S}
-p_u^2+p_{0v}^2+p_{0w}^2=-12p_0e^{3(1-\omega)u}.
\end{equation}
By using of this equation, the second equation of the system
(\ref{Q}) takes the simple form

\begin{equation}\label{T}
\dot{p_u}=3(\omega-1)p_0, \Rightarrow p_u(t)=3p_0(\omega-1)t+p_{0u},
\end{equation}where $p_{0u}$ is an integration constant which can be set to zero without loss of generality, so we set $p_{0u}=0$.

Before going to solve the rest equations of the system (\ref{Q}),
note that we may consider the relation (\ref{P}) as a gauge fixing
condition $G=T-t\approx 0$, the stability of which results the
choice $N=e^{3\omega u}$ for the lapse function. Indeed, our
dynamical system is a totally constrained classical system in which
the canonical coordinates $(\bf{q},\bf{p})$, with
${\bf{q}}=(u,v,w,T)$ and ${\bf{p}}=(p_u,p_v,p_w,P_T)$, parameterize
its kinematical phase space $\Gamma$. In such a system the classical
dynamics is given by the constraint ${\cal
C}({\bf{q},\bf{p}})\approx 0$. This means that the phase space
variables evolve on a subset of $\Gamma$, say $\bar {\Gamma}$, on
which this constraint holds. For our problem at hand this constraint
is nothing but the Hamiltonian constraint (\ref{N}). Now, to find
the gauge invariant phase space functions, i.e. Dirac observables,
we may solve the constraint  ${\cal C}({\bf{q},\bf{p}})\approx 0$.
In some special cases by this procedure the constraint takes the
form \cite{cor}

\begin{equation}\label{T1}
{\cal C}(q_i,p_i)=\bar {{\cal
C}}(q_1,...,q_{n-1};p_1,...,p_{n-1})+\alpha p_n\approx 0,
\end{equation}where $2n$ is the dimension of the phase space. For
such a constraint the momentum $p_n$ is a Dirac observable in the
sense that $\left\{p_n,{\cal C}\right\}=0$, so it is a constant of
motion. On the other hand, we note that the above constraint does
not depend on the variable $q_n$, which in turn, means that it is a
linear function of the parameter $t$, and therefore one may consider
it as a suitable clock phase space function in terms of which the
evolution of the rest phase space variables can be viewed.

Now, let us return to our problem under consideration and see how
the above procedure works. First, note that by the gauge fixing
$G=T-t\approx 0$, the Hamiltonian constraint takes the form

\begin{equation}\label{T2}
{\cal C}(u,v,w,T;p_u,p_v,p_w,P_T)=\bar {{\cal
C}}(u;p_u,p_v,p_w)+P_T,
\end{equation}where

\begin{equation}\label{T3}
\bar {{\cal
C}}(u;p_u,p_v,p_w)=\frac{1}{12}e^{-3(1-\omega)u}\left(-p_u^2+p_v^2+p_w^2\right).
\end{equation}It is seen that the constraint is of the form of the above
discussed one (\ref{T1}). So, the system de-parameterizes in the
sense that the evolution (in terms of the clock parameter $T$) of
all, but the perfect fluid degrees of freedom, can be generated by
the time independent gauge fixed Hamiltonian $\bar {{\cal C}}$
without refereing to the Hamiltonian constraint $H=0$. In this way,
it is not difficult to see the solution (\ref{T}) is again recovered
and thus to continue, we have

\begin{equation}\label{U}
\dot{u}=\frac{1}{2}(1-\omega)p_0e^{-3(1-\omega)u}t,
\end{equation}where upon integration we obtain

\begin{equation}\label{V}
u(t)=\ln \left[\frac{3}{4}p_0(1-\omega)^2
t^2+C\right]^{\frac{1}{3(1-\omega)}},
\end{equation}with $C$ being an integration constant whose value can
be evaluated in terms of the other constants by substitution of the
above expression into the constraint equation (\ref{S}) with result:
$C=-\frac{1}{12}\frac{p_{0v}^2+p_{0w}^2}{p_0}$. Therefore, the
comoving volume of the universe will be

\begin{equation}\label{W1}
V(t)=e^{3u(t)}=V_0\left(t^2-t_*^2\right)^{\frac{1}{(1-\omega)}},
\end{equation}where
$V_0=\left[\frac{3}{4}p_0(1-\omega)^2\right]^{\frac{1}{1-\omega}}$
and $t_{*}^2=\frac{p_{0v}^2+p_{0w}^2}{9p_0^2 (1-\omega)^2}$. For
this solution, the condition $V(t)\geq 0$ separates two sets of
solutions, $V_{-}(t)$ and $V_{+}(t)$, each of which is valid for
$t\leq -t^{*}$ and $t\geq t^{*}$ respectively. For the former, we
have a contracting universe which decreases its size according to a
power law relation and ends its evolution in a singularity at
$t=-t^*$, while for the latter, the evolution of the universe begins
with a big-bang singularity at $t=t^*$ and then follows the power
law expansion at late time of cosmic evolution. The corresponding
equations for the anisotropic functions in this case will be

\begin{equation}\label{X1}
\dot{v}=\frac{1}{6}p_{0v}V_0^{-(1-\omega)}\left(
t^2-t_{*}^2\right)^{-1},
\end{equation}and

\begin{equation}\label{Y1}
\dot{w}=\frac{1}{6}p_{0w}V_0^{-(1-\omega)}\left(
t^2-t_{*}^2\right)^{-1},
\end{equation}with solutions

\begin{equation}\label{Z1}
v(t)=\frac{1}{6t_{*}}p_{0v}V_0^{-(1-\omega)}\tanh
^{-1}\left(\frac{t}{t_*}\right),\hspace{5mm}w(t)=\frac{1}{6t_{*}}p_{0w}V_0^{-(1-\omega)}\tanh
^{-1}\left(\frac{t}{t_*}\right).
\end{equation}
It is seen that, unlike the volume factor which was not defined in
the time interval $(-t_*,t_*)$, the functions $v(t)$ and $w(t)$ are
only well-defined in this interval. Therefore, in general we do not
get acceptable physical solutions. However, for some special values
of EoS parameter, say for $\omega=\frac{2n-1}{2n}, n=1,2,...$, the
volume takes the form $V(t)=V_0\left(t^2-t_*^2\right)^{2n}$ which is
always positive and well-defined also in $(-t_*,t_*)$. For such
special occasions, we are facing with a universe that begins its
evolution from a singular point at $t=-t_*$ with zero size and high
degrees of anisotropy, expands to a isotropic state with maximum
volume and then will collapse and eventually end its evolution in a
singularity similar to its initial singularity.

At the end of this section let us see what happened if $C=0$, for
which we get $p_{0v}=p_{0w}=0$. In this case the two last equations
of the system (\ref{Q}) give $\dot{v}=\dot{w}=0$. Therefore, we have

\begin{equation}\label{W2}
V(t)=V_0t^{\frac{2}{(1-\omega)}},
\hspace{5mm}v(t)=v_0=\mbox{const},\hspace{5mm}w(t)=w_0=\mbox{const}.
\end{equation}In this situation, a look at the scale factors in (\ref{B}) shows that
they take the form

\begin{equation}\label{W3}
a(t)=a_0e^{u(t)},\hspace{5mm}b(t)=b_0e^{u(t)},\hspace{5mm}c(t)=c_0e^{u(t)},
\end{equation}where $a_0=e^{v_0+\sqrt{3}w_0}$,
$b_0=e^{v_0-\sqrt{3}w_0}$ and $c_0=e^{-2v_0}$. Under these
conditions, by re-definition of the coordinates as $X=a_0x$,
$Y=b_0y$ and $Z=c_0z$, the metric will be

\begin{equation}\label{W4}
ds^2=-N^2(t)dt^2+e^{2u(t)}\left(dX^2+dY^2+dZ^2\right),
\end{equation}which is nothing but the flat FRW metric. Since expression (\ref{W2}) for $V(t)$ is singular at $t=0$, we may
consider the intervals $-\infty<t\leq 0$ or $0\leq t<+\infty$ as the
domain of variation for the time parameter $t$. Thus, with $0\leq
t<+\infty$, the above solutions give a flat FRW universe begins to
evolve from a big-bang singularity and continually expands according
to the power law expression giving in (\ref{W2}).

In the next section we will deal with the quantization of the model
described above to see how the presented classical picture can be
modified.

\section{Quantization of the model}
In the previous section we saw that classical model consisted of two
categories of solutions (correspond to the cases $C\neq 0$ and
$C=0$), which in general, suffered from the existence of big-bang
singularities. In this section we focus our attention on the
quantization of the model described above. Quantum cosmology of the
Bianchi models are widely investigated in literature, see for
instance \cite{QBianchi} and the references therein. Our starting
point is the WDW equation $\hat{H}\Psi(u,v,w,T)=0$, in which
$\hat{H}$ is the operator version of the Hamiltonian (\ref{N}) and
$\Psi(u,v,w,T)$ is the wave function of the universe. The operator
$\hat{H}$ can be constructed by replacing $\hat{q}\Psi=q\Psi$ and
$\hat{p_q}\Psi=-i\frac{\partial}{\partial q}\Psi$, for each
dynamical variable. Nevertheless, a point to be considered is the
issue of the factor-ordering problem when one is going to arrange a
quantum mechanical operator equation. In dealing with such
Hamiltonians at the quantum level one should be careful when trying
to replace the dynamical variables with their quantum operator
counterparts, that is, in replacing a variable $q$ and its momentum
$p_q$ with their corresponding operators, the ordering
considerations should be taken into account. In our present case,
this is only important in the first term of the Hamiltonian, since
this term includes variables $u$ and $p_u$ that do not commute.
Taking these considerations into account, the WDW equation for the
Hamiltonian (\ref{N}) is written as

\begin{equation}\label{AC}
\left[2\frac{\partial^2}{\partial u^2}+3q\frac{\partial}{\partial
u}-2\frac{\partial^2}{\partial v^2}-2\frac{\partial^2}{\partial
w^2}-24ie^{3(1-\omega)u}\frac{\partial}{\partial
T}\right]\Psi(u,v,w,T)=0,
\end{equation}where $q$ is the factor ordering parameter which represents the ambiguity in the
ordering of factors $u$ and $p_u$ in the first term of (\ref{N}). It
is clear that there are infinite number of possibilities to choose
this parameter. For example $q=0$ corresponds to no factor ordering,
with $q=1$ the kinetic term of the Hamiltonian takes the form of the
Laplacian $-\frac{1}{2}\nabla^2$ of the minisuperspace and $q=-2$
corresponds to the ordering $e^{-3u}p_u^2=p_ue^{-3u}p_u$. Although
in general, the behavior of the wavefunction depends on the chosen
factor ordering \cite{factor}, it can be shown that the
factor-ordering parameter will not affect semiclassical calculations
in quantum cosmology \cite{Haw}, and so for convenience one usually
chooses a special value for it in the special models. With an eye to
the constraint equation (\ref{T2}), we can see that equation
(\ref{AC}) is nothing other than ${\cal C}\Psi=\left(\bar{{\cal
C}}+P_T\right)\Psi=0$. In our classical analysis, we saw that the
momentum of the perfect fluid $P_T$, is a constant of motion and
hence we have taken its value to be $P_T=p_0=\mbox{const}$. On the
other hand our discussion about the sign of the volume function in
equation (\ref{W1}) shows that the sign of $p_0$ should, in general,
be positive. Therefore, classically, the motion of the system is
restricted to those regions of the phase space for which we have
$\bar {{\cal
C}}=\frac{1}{12}e^{-3(1-\omega)u}\left(-p_u^2+p_v^2+p_w^2\right)<0$.
Now, with separation ansatz

\begin{equation}\label{AD}
\Psi(u,v,w,T)=e^{iET}\psi(u,v,w),
\end{equation}for the quantum version of the constraint, we obtain

\begin{equation}\label{AE}
\left[2\frac{\partial^2}{\partial u^2}+3q\frac{\partial}{\partial
u}-2\frac{\partial^2}{\partial v^2}-2\frac{\partial^2}{\partial
w^2}+24Ee^{3(1-\omega)u}\right]\psi(u,v,w)=0.
\end{equation}Comparing of this equation with the classical
constraint ${\cal C}=\bar{{\cal C}}+P_T\approx 0$, shows that the
separation constant $E$ plays the role of the classically constant
$P_T$ and thus when constructing wave packets a superposition over
non-negative values of $E$ should be taken (see equation (\ref{AJ})
below). Note that this choice for the sign of $E$ has in particular
the consequence, that the Hamiltonian involved in the WDW equation
is given by the quantization of the classical expression $|\bar{\cal
C}|$, as can be seen directly from equation (\ref{AC}) as well as
from the fact that $\bar{\cal C}$ is negative in the classical
theory. Separation of variables in the form
$\psi(u,v,w)=U(u)V(v)W(w)$ yields the following form for the
functions $V(v)$ and $W(w)$:

\begin{equation}\label{AF}
V(v)=e^{\pm k_1v},\hspace{5mm}W(w)=e^{\pm k_2w},
\end{equation}where $k_1$ and $k_2$ are separation constants where without loss of generality, we can choose positive values for them.
At this point, let us take a look at the boundary condition on the
$V(v)$ and $W(w)$ sector of the wave function. If we assume that for
infinite values for the variables $v$ and $w$, the wave function
will be zero, we may take the solutions (\ref{AF}) with upper
(lower) sign in the exponential function for $v,w<0$ ($v,w>0$). This
means that the function $V(v)$ and $W(w)$ can be written as

\begin{equation}\label{AF1}
V(v)=e^{-k_1|v|},\hspace{5mm}W(w)=e^{-k_2|w|}.
\end{equation}
Also, for $U(u)$ we get

\begin{equation}\label{AG}
\left[2\frac{d^2}{du^2}+3q\frac{d}{du}+\left(24Ee^{3(1-\omega)u}-\nu^2\right)\right]U(u)=0,
\end{equation}where $\nu^2=2(k_1^2+k_2^2)$. The solution of the
above equation may be expressed in terms of the Bessel functions $J$
and $Y$. However, for having well-defined functions in all ranges of
the arguments of the Bessel functions, we only keep the function
$J$, in terms of which the solution is

\begin{equation}\label{AH}
U(u)=e^{-\frac{3}{4}qu}J_{\frac{\sqrt{9 q^2+8 \nu ^2}}{6(1-\omega)
}}\left(\frac{4 \sqrt{E} e^{\frac{3}{2} (1-\omega) u}}{\sqrt{3}
(1-\omega)}\right).
\end{equation}Therefore, the eigenfunctions of the WDW equation may
be written as

\begin{equation}\label{AI}
\Psi_{E,\nu}(u,v,w,T)=e^{-\frac{3}{4}qu}e^{iET}e^{-(|v|+|w|)\nu}J_{\frac{\sqrt{9
q^2+8 \nu ^2}}{6(1-\omega) }}\left(\frac{4 \sqrt{E} e^{\frac{3}{2}
(1-\omega) u}}{\sqrt{3} (1-\omega)}\right),
\end{equation}in which since the behavior of $v$ and $w$ is similar
we have taken $k_1=k_2=\nu/2$ and also the variables $v,w$ are
re-scaled as $|v|/2\rightarrow |v|, |w|/2\rightarrow |w|$. The wave
function of the universe is indeed the general solution of the WDW
equation which can be made from the superposition of its
eigenfunctions. Thus, we may write the wave function as

\begin{equation}\label{AJ}
\Psi(u,v,w,T)=\int_{E=0}^\infty
\int_{\nu=0}^{\infty}A(E)C(\nu)\Psi_{E,\nu}(u,v,w,T)dEd\nu,
\end{equation}where $A(E)$ and $C(\nu)$ are some weight functions
with the help of which suitable wavepacket will be constructed.
Indeed, the wave function in quantum cosmology should be constructed
in such a way as to achieve an acceptable match to the classical
model. This means that by the above relation, one usually constructs
a coherent wavepacket with good asymptotic behavior in the
minisuperspace, peaking in the vicinity of the classical trajectory.
So, with the help of equality \cite{Hand}

\begin{equation}\label{AK}
\int_0^\infty
e^{-ar^2}r^{\eta+1}J_{\eta}(br)dr=\frac{b^{\eta}}{(2a)^{\eta+1}}e^{-\frac{b^2}{4a}},
\end{equation}the integral over $E$ may be evaluated to yield analytical expression if we choose the weight factor $A(E)$ to be a quasi-Gaussian function as

\begin{equation}\label{AL}
A(r)=\frac{8}{3(1-\omega)^2}r^{\eta}e^{-\gamma r^2},
\end{equation}where $\gamma$ is a positive constant and
$r=\frac{4\sqrt{E}}{\sqrt{3}(1-\omega)}$. With this, integration
over $E$ will be done and we arrive at the following expression for
the wave function

\begin{equation}\label{AM}
\Psi(u,v,w,T)=\frac{1}{2a}e^{-\frac{3}{4}qu}e^{-\frac{b^2}{4a}}\int_{\nu=0}^{\infty}C(\nu)e^{-(|v|+|w|)\nu}\left(\frac{b}{2a}\right)^{\eta}d\nu,
\end{equation}
where $b=e^{\frac{3}{2}(1-\omega)u}$,
$a=\gamma-\frac{3}{16}(1-\omega)^2iT$ and
$\eta=\frac{\sqrt{9q^2+8\nu^2}}{6(1-\omega)}$. A point to be noted
is that our choices for $A(E)$ as quasi-Gaussian function also has
physical grounds since these types of functions are widely used in
quantum mechanics to build localized states. The reason is that
these functions have a peak around a particular point of their
argument and rapidly decrease as they move away from that point.
This makes the wavepacket generated by the above relation (after
integration over $\nu$), behaves the same way, i.e. is localized
around some special values of its arguments.

Now, let us return to the superposition over $\nu$ which as seen
from (\ref{AM}) finding a function $C(\nu)$ that leads to an
analytical expression for the wave function is not an easy task.
However, notice that the integrand in (\ref{AM}) only has a
significant amount for the small values of $\nu$ and becomes smaller
and smaller as $\nu$ increases. Under this condition, we may take
the wave function as being represented by relation (\ref{AM}) with
the integral to be truncated at a suitable value of $\nu$ displaying
this property. So, if we assume that the superposition over $\nu$ is
taken over some values of $\nu$ in the range $\nu<<q$, we have $\eta
\simeq \frac{q}{2(1-\omega)}$ and the integral (\ref{AM}) takes the
form

\begin{equation}\label{AN}
\Psi(u,v,w,T)=\frac{1}{(2a)^{\eta+1}}e^{-\frac{b^2}{4a}}\int_{\nu=0}^{\sigma}C(\nu)e^{-(|v|+|w|)\nu}d\nu,
\end{equation}where $\sigma$ is a positive constant. Now, we may use the equality \cite{Hand}

\begin{equation}\label{AO}
\int_0^\sigma
\nu^{\alpha-1}e^{-p\nu}d\nu=p^{-\alpha}\gamma(\alpha,\sigma
p),\hspace{5mm}\mbox{Re} (\alpha)>0,
\end{equation}where $\gamma(a,x)$ is the incomplete Gamma function
\[\gamma(\alpha,z)=\int_0^ze^{-t}t^{\alpha-1}dt,\hspace{5mm}\mbox{Re}(\alpha)>0,\]
to take the weight factor $C(\nu)$ as $C(\nu)=\nu^{\alpha-1}$. At
the first look, it seems that this form for the function $C(\nu)$
(which may look rather ad hoc) is chosen in such a way that yields
an analytical expression for the wave function. However, note that
for small values of $\nu$, we have $\mu^2=1-\nu^{\alpha-1}>0$, so we
may write $e^{-\mu^2}=e^{-(1-\nu^{\alpha-1})}\simeq
1-\mu^2=\nu^{\alpha-1}$. This shows that the in the domain of $\nu$,
for which we are going to perform the superposition (\ref{AN}), the
function $C(\nu)=\nu^{\alpha-1}$ may be approximated with a Gaussian
function whose physical grounds are described above. This lead us to

\begin{equation}\label{AP}
\Psi(u,v,w,T)=\frac{1}{(2a)^{\eta+1}}e^{-\frac{b^2}{4a}}\left(|v|+|w|\right)^{-\alpha}\gamma\left(\alpha,\sigma
(|v|+|w|)\right).
\end{equation}Finally, in terms of the volume $V=e^{3u}$ the wave
function reads as (we rescale $\frac{3}{16}T\rightarrow t$ and
$\frac{V^{1-\omega}}{4}\rightarrow V^{1-\omega}$)

\begin{equation}\label{AQ}
\Psi(V,v,w,t)=\frac{{\cal
N}}{\left[\gamma-(1-\omega)^2it\right]^{\eta+1}} \exp
\left[-\frac{V^{1-\omega}}{\gamma-(1-\omega)^2it}\right]
\left(|v|+|w|\right)^{-\alpha}\gamma\left(\alpha,\sigma(|v|+|w|)\right),
\end{equation}where ${\cal N}$ is a normalization factor. Now, with this
wave function we are interested to see how the evolution of the
dynamical variables is predicted in the framework of the quantum
model. In fact, what we expect from a consistent model of quantum
cosmology is that its late time predictions are in agreement with
the classical model, but should be separated from classical
solutions in the early times of cosmic evolution where classical
singularities occur. To see how this procedure works in our model,
let us evaluate the time dependence of the expectation value of a
dynamical variable $q$ as

\begin{equation}\label{AQ1}
<q>(t)=\frac{<\Psi|q|\Psi>}{<\Psi|\Psi>}.
\end{equation}To calculate the expectation values, note that the WDW
equation (\ref{AC}) is like a Schr\"{o}dinger equation $i\partial
\Psi/\partial T=H\Psi$, in which the Hamiltonian operator is
Hermitian with respect to the inner product

\begin{equation}\label{AQ2}
<\Phi|\Psi>=\int_{(V,v,w)}V^{1-\omega}\Phi^*\Psi dV dv dw.
\end{equation}Now, we may write the the expectation value for
the volume as

\begin{equation}\label{AR}
<V>(t)=\frac{\int_{V=0}^{\infty}\int_{v=-\infty}^{\infty}\int_{w=-\infty}^{\infty}V^{1-\omega}\Psi^*V\Psi
dVdvdw}
{\int_{V=0}^{\infty}\int_{v=-\infty}^{\infty}\int_{w=-\infty}^{\infty}V^{1-\omega}\Psi^*\Psi
dVdvdw},
\end{equation}which yields

\begin{equation}\label{AS}
<V>(t)=V_0\left[\gamma ^2+
(1-\omega)^4t^2\right]^{\frac{1}{1-\omega}}.
\end{equation}
Also, the expectation values of the anisotropic functions read as

\begin{equation}\label{AT}
<v>(t)=\frac{\int_{V=0}^{\infty}\int_{v=-\infty}^{\infty}\int_{w=-\infty}^{\infty}V^{1-\omega}\Psi^*v\Psi
dVdvdw}
{\int_{V=0}^{\infty}\int_{v=-\infty}^{\infty}\int_{w=-\infty}^{\infty}V^{1-\omega}\Psi^*\Psi
dVdvdw},\hspace{5mm}\mbox{similar expression for $<w>(t)$,}
\end{equation}
with the result

\begin{equation}\label{AU}
<v>(t)=\frac{\int_{v,w=-\infty}^{\infty}v\left[\left(|v|+|w|\right)^{-\alpha}\gamma\left(\alpha,\sigma(|v|+|w|)\right)\right]^2dvdw}
{\int_{v,w=-\infty}^{\infty}\left[\left(|v|+|w|\right)^{-\alpha}\gamma\left(\alpha,\sigma(|v|+|w|)\right)\right]^2dvdw}=0,\hspace{5mm}\mbox{similarly:}
\hspace{2mm}<w>(t)=0.
\end{equation}
It is important whether the quantum solutions are capable of solving
the singularity of the classical models. We see that the quantum
wave function (\ref{AQ}) gives the evolution of the expectation
value of the volume factor as a nonsingular (never vanishing)
bouncing function (\ref{AS}), which its late time behavior coincides
to the late time behavior of the classical solution (\ref{W1}) and
(\ref{W2}), that is $V(t)\sim t^{\frac{2}{1-\omega}}$. Another
feature of the quantum picture is that it also predicts a zero
expectation value for the anisotropy of the corresponding universe.
This means that unlike classical models in which the expansion of
the universe was anisotropic (and in some cases even with infinite
amounts), the quantization of the system predicts an isotropic
expansion. In summary, as a result of the quantization of this
cosmological model, the two classical solutions presented in the
previous section are united in a single non-singular picture.

\section{Bohmian trajectories}
In the previous section, we saw that quantization of the model based
on the canonical quantization procedure led to the replacement of
the classical singular solutions with a non-singular bouncing
picture. Now, we might ask what causes the bounce? It is clear that
the answer to this question should be sought in the emergence of
quantum effects that show themselves when the size of the universe
is very small. To deal with this question, we are going to use the
ontological (Bohmian) interpretation of quantum mechanics
\cite{Bohm}. In this formulation of the quantum mechanics one writes
the wave function in the polar form $\Psi({\bf q},t)=\Omega({\bf
q},t) e^{iS({\bf q},t)}$, where $\Omega$ and $S$ are some real
functions of the configuration space variables and time. For the
wavefunction (\ref{AQ}) these function take the form

\begin{equation}\label{AV}
\Omega(V,v,w,t)=\left[\gamma^2+(1-\omega)^4t^2\right]^{-\frac{\eta+1}{2}}
\exp\left(-\frac{\gamma}{\gamma^2+(1-\omega)^4t^2}V^{1-\omega}\right)f(v,w),
\end{equation}

\begin{equation}\label{AX}
S(V,v,w,t)=(\eta+1)\arctan
\frac{(1-\omega)^2t}{\gamma}-\frac{(1-\omega)^2t}{\gamma^2+(1-\omega)^4t^2}V^{1-\omega},
\end{equation}where
$f(v,w)=\left(|v|+|w|\right)^{-\alpha}\gamma\left(\alpha,\sigma(|v|+|w|)\right)$.
In the Bohm-de Broglie interpretation of quantum mechanics the
central equations come from substituting the above mentioned polar
form of the wave function into the Schr\"{o}dinger (or WDW)
equation. By this procedure one gets a continuity equation and the
modified Hamilton-Jacobi equation as, (see \cite{Bohm1} for details)

\begin{equation}\label{AY}
H\left(q_i,p_i=\frac{\partial S}{\partial q_i}\right)+{\cal Q}=0,
\end{equation}where ${\cal Q}$ is the quantum potential which in our
case a simple algebra gives it as

\begin{equation}\label{AZ}
{\cal
Q}=\frac{1}{V^{1-\omega}\Omega}\left[18V^2\frac{\partial^2\Omega}{\partial
V^2}+9(q+2)V\frac{\partial \Omega}{\partial
V}-2\left(\frac{\partial^2\Omega}{\partial
v^2}+\frac{\partial^2\Omega}{\partial w^2}\right)\right].
\end{equation}The dynamical behavior of the variables will be
determined from the key equation $p_i=\frac{\partial S}{\partial
q_i}$. To see how this mechanism works, first note that with
$N=e^{3\omega u}=V^{\omega}$ the Lagrangian (\ref{J}) takes the form
${\cal L}_g=-\frac{1}{3}\frac{\dot{V}^2}{V^{\omega+1}}+...$, from
which we obtain $p_V=-\frac{2}{3}\frac{\dot{V}}{V^{\omega+1}}$.
Therefore, from (\ref{AX}), the equation $p_V=\frac{\partial
S}{\partial V}$ gives us

\begin{equation}\label{BA}
\frac{2}{3}\frac{\dot{V}}{V^{\omega+1}}=\frac{(1-\omega)^3t}{\gamma^2+(1-\omega)^4t^2}V^{-\omega},
\end{equation}from which, after integration we obtain the Bohmian representation
of the volume factor as (considering that we have already re-scaled
$t$ and $V$ when writing the wavefunction (\ref{AQ}))

\begin{equation}\label{BC}
V(t)\sim \left[\gamma^2+(1-\omega)^4t^2\right]^{\frac{1}{1-\omega}}.
\end{equation}Also, since $\frac{\partial S}{\partial v}=\frac{\partial S}{\partial
w}=0$, from (\ref{K}) we arrive at

\begin{equation}\label{BD}
\dot{v}=\dot{w}=0\Rightarrow v,w=\mbox{const.}
\end{equation}These expressions have exactly the same dynamical behavior that we had previously obtained from the quantum cosmological
model. As is clear, here too, the bouncing (from a non-zero value)
behavior near the classical singularity is the main property of the
volume function and the anisotropy functions such as the quantum
model have a constant value. A point to be emphasized here is that,
in addition to preventing the classical singularities, the
appearance of a bounce in the quantum and Bohemian models, from this
aspect is also interesting that it predicts the existence of a
minimum size when the universe is evolving. This is important
because we know that the existence of a minimal length in nature is
an idea supported by all candidates of quantum gravity.

\begin{figure}
\includegraphics[width=2in]{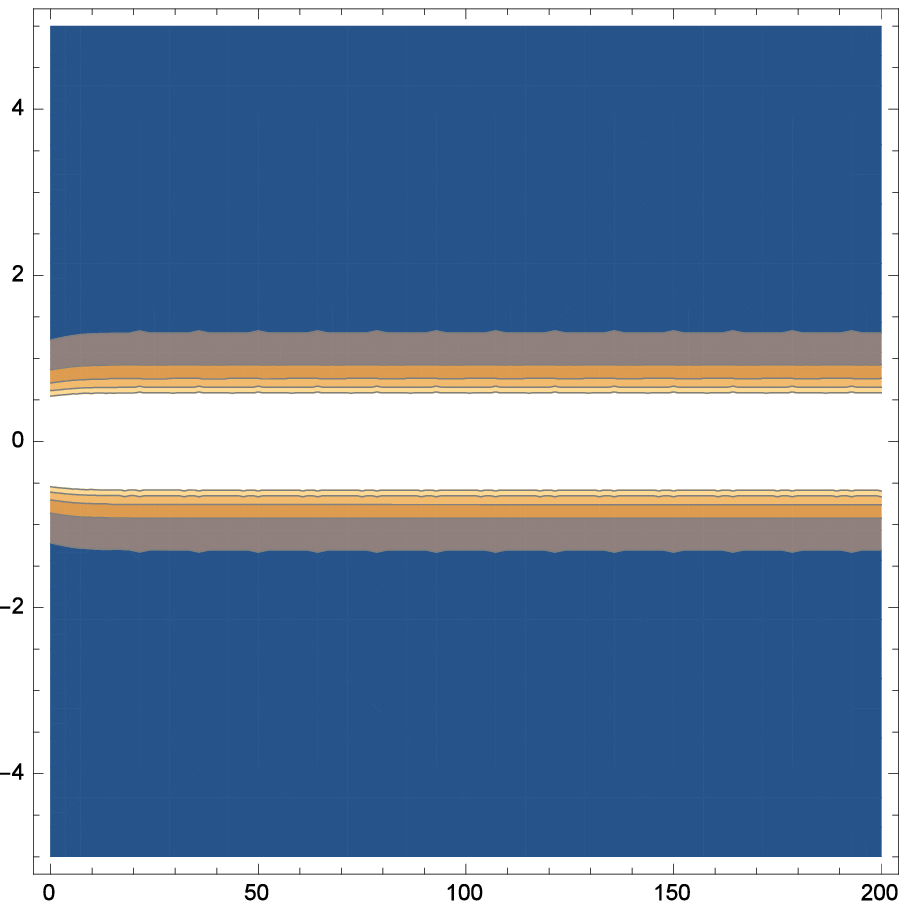}\hspace{50mm}\includegraphics[width=2in]{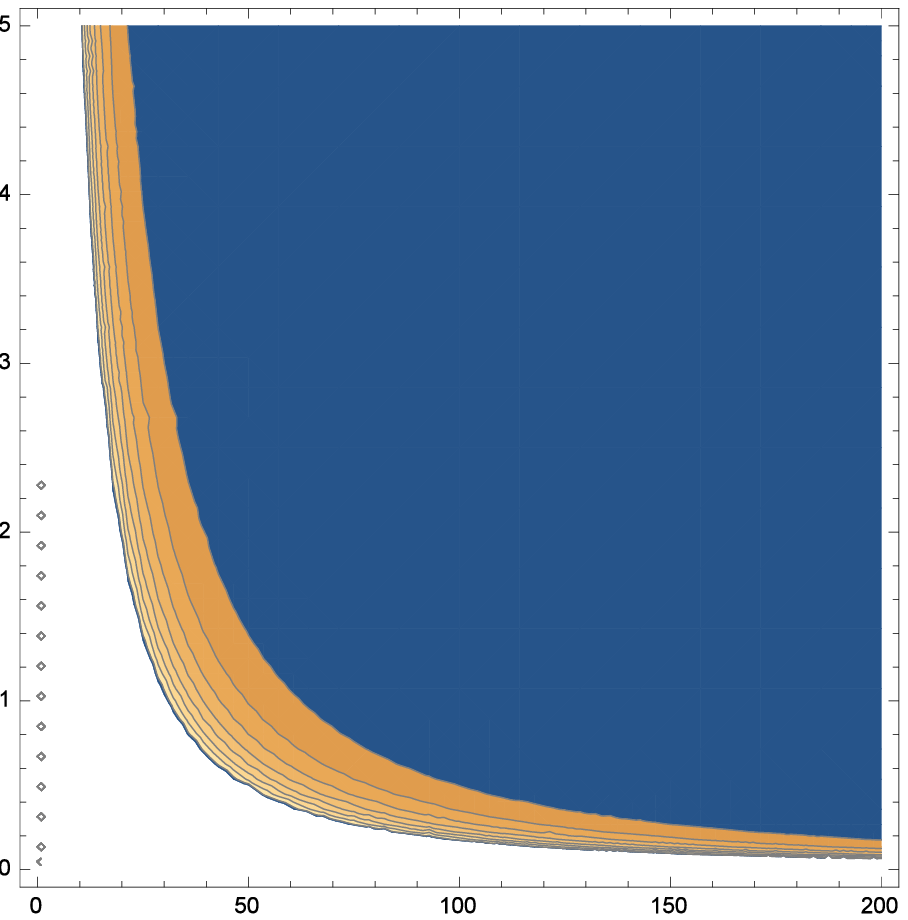}\\
\caption{The contour plots of ${\cal F}_V$ (left) and ${\cal
F}_{(v,w)}$ (right). The figures are plotted for the numerical
values: $\omega=-1$, $q=1$, $\sigma=1$ and $\alpha=2$. As the
figures show, these quantities get their significant amounts (white
area) near the classical singularity. For ${\cal F}_V$ (left), to
better see this issue, we have expanded the $V$-axis to negative
values, although we know that $V$ is positive.}\label{fig1}
\end{figure}

To understand the origin of the singularity avoidance, let us
evaluate the quantum potential in terms of the volume and
anisotropic factors. With the help of the equation (\ref{BC}) the
function $\Omega$ takes the form

\begin{equation}\label{BE}
\Omega(V,v,w)=V^{-\frac{1}{4} (q-2 \omega
+2)}\left(|v|+|w|\right)^{-\alpha}\gamma\left(\alpha,\sigma(|v|+|w|)\right),
\end{equation}by means of which and equation (\ref{AZ}) we arrive at
the following expression for the quantum potential

\begin{eqnarray}\label{BF}
{\cal Q}(V,v,w)&=&-\frac{9}{8}(q-2 \omega +2) (q+2 \omega -2)
V^{\omega-1} \\ \nonumber &+& V^{\omega-1}\frac{e^{-\sigma (v+w)}
\left[4(\sigma(v+w))^{\alpha}(\alpha+\sigma (v+w)+1)-4 \alpha
(\alpha+1) e^{\sigma (v+w)}
\gamma\left(\alpha,\sigma(v+w)\right)\right]}{(v+w)^2
\gamma\left(\alpha,\sigma(v+w)\right)}.
\end{eqnarray}
As this relation shows, the quantum potential goes to zero when the
volume function is large (note that $\omega-1<0$) which is an
expected behavior since in this regime the universe evolves
classically and so the  quantum effects can be ignored. On the other
hand, by reducing the volume function, the quantum potential
increases and here is where the quantum effects are important in
cosmology. Indeed, in this regime a huge repulsive force may be
produced which can be extracted from the quantum potential as
$\overrightarrow{{\cal F}}=-\nabla {\cal Q}$, with components

\begin{equation}\label{BG}
{\cal F}_V=-\frac{\partial {\cal Q}}{\partial V}=-
\frac{(9/8)(1-\omega) (q-2 \omega +2) (q+2 \omega
-2)-(1-\omega)h(v,w)} {V^{2-\omega}},
\end{equation}

\begin{equation}\label{BH}
{\cal F}_{(v,w)}=-\frac{\partial {\cal Q}}{\partial
(v,w)}=-\frac{e^{-2\sigma(v+w)}}{V^{1-\omega}}g(v,w),
\end{equation}where $h(v,w)$ is the fraction appearing in the second
term of (\ref{BF}) and $g(v,w)=\partial_v h/e^{-2\sigma(v+w)}$. From
the above expressions it is clear that in the vicinity of classical
singularity (for small values of $V$) this force becomes large and
its repulsive nature prevents the volume to evolve to zero size but
instead a bounce will occur. In summary, the above discussion shows
that quantum potential and consequently quantum effects are
important when the universe is in its early stages and in the late
time of cosmic evolution, i.e. in the limit of the large scale
factors these effects can be neglected, see figure \ref{fig1}.
Therefore, asymptotically the classical behavior is recovered.

\section{Summary}
In this paper we have studied the Bianchi type I cosmological model
with a perfect fluid as its matter content. To write the Hamiltonian
of the corresponding dynamical system we have used the Schutz'
formalism for perfect fluid, the advantages of which is that it
allows us to introduce a time parameter in terms of the one of the
fluid's thermodynamical variable. Indeed, since the momentum
conjugate to this variable appears linearly in the fluid's
Hamiltonian, this formalism helps one to have an eye to the problem
of time when quantizing the system.

Based on this time gauge, we constructed the Hamiltonian equations
of motion, solving of them led us to two classes of classical
solution. The first one gave us unacceptable physical solutions
except for special values of EoS parameter in the form
$\omega=\frac{2n-1}{2n}$. The evolution of volume factor in this
case is such that, from a big-bang singularity it begins an power
law expansion until reaches a maximum value and then its contraction
phase starts and ends in a singularity. The evolution of the
universe in this case is anisotropic with an infinite of anisotropy
in the singular points. Finally, by another set of classical
solutions, we obtained an expanding volume factor with singularity
at $t=0$ in which the volume of the universe is zero. While the
universe expands its anisotropy remains constant which by a
re-scaling of the coordinates we showed that this case can be viewed
as a flat FRW universe.

We then have dealt with quantization of the cosmological setting.
Here, because of our specific choice of time gauge, the WDW equation
took the form of a Schr\"{o}dinger equation allowed us to have a
time dependent wave function. We showed that the WDW equation can be
separated and its eigenfunctions can be obtained in terms of known
special functions. By finding suitable weight function we
constructed the superposition of the eigenfunctions which led us to
an appropriate wave function. By means of the obtained wave function
we got the expectation values of the volume and anisotropic
functions. Based on the results, we verified that while the quantum
solutions recover the late time behavior of classical models, they
predict a non-singular bouncing universe in the early times of
cosmic evolutions. Therefore, the avoidance of classical
singularities and the recovery of the late time classical dynamics
are important features of our quantum model.

In the last part of the article, we returned to the quantum picture
by using of the Bohmian approach to quantum mechanics. It is shown
that the resulting expressions for the dynamical variables of the
system are the same as those previously obtained from the canonical
quantization of the model. However, the use of the Bohmian
trajectories has helped us to understand the origin of the avoidance
of singularity by a repulsive force due to the existence of the
quantum potential. We saw that near the classical singularities the
quantum potential gets a huge amount and thus the repulsive nature
of its corresponding force prevents the universe to reach the
singularity. This is while this force tends to zero at late times
which means that the quantum effects are negligible there and the
universe behaves classically.


\begin{thebibliography}{1}
\bibitem{DeW} B. S. DeWitt, {\it Phys. Rev.} {\bf 160} (1967) 1113

\bibitem{lqg}C. Rovelli and L. Smolin, {\it Nucl. Phys.} B {\bf 442} (1995)
593\\A. Ashtekar and J. Lewandowski, {\it Class. Quantum Grav.} {\bf
A14} (1997) A55\\C. Rovelli, {\it Living Rev. Relativ.} {\bf 1}
(1998) 1

\bibitem{Wilt}J. J. Halliwell, {\it Introductory Lectures on
Quantum Cosmology} (arXiv: 0909.2566 [gr-qc])\\C. Kiefer, {\it
Quantum Gravity} (Oxford University Press, New York, 2007)\\M.
Bojowald, {\it  Living Rev. Relativ.} {\bf 8} (2005) 11 (arXiv:
gr-qc/0601085)


\bibitem{Ryan}M. P. Ryan, {\it Hamiltonian Cosmology} (Springer, Berlin, 1972)\\J. Wainwright, {\it Gen. Rel.
Grav.} {\bf 16} (1984) 657\\S. Hervik, {\it Class. Quantum Grav.}
{\bf 17} (2000) 2765 (arXiv: gr-qc/0003084)\\V. N. Folomeev and V.
T. Gourovich, {\it Grav. Cosmol.} {\bf 6} (2000) 19 (arXiv:
gr-qc/0001065)\\T. Christodoulakis, T. Gakis and G. O. Papadopoulos,
{\it Class. Quantum Grav.} {\bf 19} (2002) 1013 (arXiv:
gr-qc/0106065)\\B. Vakili, N. Khosravi and H. R. Sepangi, {\it
Class. Quantum Grav.} {\bf 24} (2007) 931 (arXiv: gr-qc/0701075)\\B.
Vakili and N. Khosravi, {\it Phys. Rev.} D
{\bf 82} (2010) 103509 (arXiv: 1010.1933 [gr-qc])\\

\bibitem{Schutz}B. F. Schutz, {\it Phys. Rev.} D {\bf 2} (1970)
2762\\B. F. Schutz, {\it Phys. Rev.} D {\bf 4} (1971) 3559

\bibitem{Time}A. B. Batista, J. C. Fabris, S. V. B. Goncalves and J. Tossa, {\it Phys. Rev.} D {\bf 65} (2002) 063519
(arXiv: gr-qc/0108053)\\P. Pedram and S. Jalalzadeh, {\it Phys.
Rev.} D {\bf 77} (2008) 123529 (arXiv: 0805.4099 [grqc])\\B. Vakili,
{\it Phys. Lett.} B {\bf 688} (2010) 129 (arXiv: 1004.0306
[gr-qc])\\B. Vakili, {\it Class. Quantum Grav.} {\bf 27} (2010)
025008 (arXiv: 0908.0998 [gr-qc])

\bibitem{Misner}C. W. Misner, {\it Phys. Rev.} {\bf 186} (1969) 1319

\bibitem{book1}{\O}. Gr{\o}n and S. Hervik, {\it Einsteins General Theory of Relativity:
With Modern Applications in Cosmology} (Springer-Verlag, New York,
NY, USA, 2007)

\bibitem{cor}A. Corichi and T. Vuka\v{s}inac, {\it Phys. Rev.} D {\bf 86} (2012) 064019 (arXiv: 1202.1846
[gr-qc])\\A. Corichi and T. Vuka\v{s}inac, {\it On the choice of
time in the continuum limit of polymeric effective theories} (arXiv:
1209.2752 [gr-qc])

\bibitem{QBianchi}F. G. Alvarenga, A. B. Batista, J. C. Fabris and S. V. B. Gon\c{c}alves, {\it Gen. Rel. Grav.} {\bf 35} (2003)
1659 (arXiv: gr-qc/0304078)
\\J. Malecki, {\it  Phys. Rev.} D {\bf 70} (2004)
084040 (arXiv: gr-qc/0407114)\\G. Date, {\it Phys. Rev.} D {\bf 72}
(2005) 067301 (arXiv: gr-qc/0505030)\\B. Vakili and H. R. Sepangi,
{\it Phys. Lett.} B {\bf 651} (2007) 79 (arXiv: 0706.0273
[gr-qc])\\A. Ashtekar and E. Wilson-Ewing, {\it Phys. Rev.} D {\bf
79} (2009) 083535 (arXiv: 0903.3397 [gr-qc])\\S. Pal and N.
Banerjee, {\it Phys. Rev.} D {\bf 90} (2014) 104001 (arXiv:
1410.2718 [gr-qc])\\R. Moriconi, G. Montani and S. Capozziello, {\it
Phys. Rev.} D {\bf 94} (2016) 023519 (arXiv: 1607.04481 [gr-qc])\\A.
Karagiorgos, T. Pailas, N. Dimakis, P. A. Terzis and T.
Christodoulakis, {\it JCAP} {\bf 03} (2018) 030 (arXiv: 1710.02032
[gr-qc])\\S. Ghosh, S. Gangopadhyay and P. K. Panigrahi, {\it Mod.
Phys. Lett.} A {\bf 34} (2019) 1950283 (arXiv: 1809.03312 [gr-qc])

\bibitem{factor}R. Steigl and F. Hinterleitner, {\it Class. Quantum Grav.} {\bf 23} (2006) 3879

\bibitem{Haw}S. W. Hawking and D. N. Page, {\it Nucl. Phys.} B {\bf 264} (1986) 185

\bibitem{Hand}I. S. Gradshteyn, I. M. Ryzhik, A. Jeffrey and D. Zwillinger, {\it Table of Integrals, Series and Products} (Academic Press, San Diego,
2007)

\bibitem{Bohm}D. Bohm, {\it Phys. Rev.} {\bf 85} (1952) 166\\
D. Bohm, {\it Phys. Rev.} {\bf 85} (1952) 180\\ P. R. Holland, {\it
The Quantum Theory of Motion: An Account of the de Broglie-Bohm
Interpretation of Quantum Mechanics} (Cambridge University Press,
Cambridge, 1993)

\bibitem{Bohm1}F. T. Falciano and N. Pinto-Neto, {\it Phys. Rev.} D {\bf 79} (2009) 023507\\A. Shojai and F. Shojai, {\it Europhys. Lett.} {\bf 71} (2005) 886
\end{thebibliography}
\end{document}